\documentclass[11pt,a4paper]{article}
\pdfoutput=1
\usepackage{jheppub}

\newsavebox{\uuunit}
\sbox{\uuunit}
    {\setlength{\unitlength}{0.825em}
     \begin{picture}(0.6,0.7)
        \thinlines
        \put(0,0){\line(1,0){0.5}}
        \put(0.15,0){\line(0,1){0.7}}
        \put(0.35,0){\line(0,1){0.8}}
       \multiput(0.3,0.8)(-0.04,-0.02){12}{\rule{0.5pt}{0.5pt}}
     \end {picture}}
\newcommand {\unity}{\mathord{\!\usebox{\uuunit}}}

\newcommand{\be}{\begin{equation}}
\newcommand{\ee}{\end{equation}}

\newcommand{\ffour}{{\cal F}^{\parallel}}
\newcommand{\ffourtwo}{F}
\newcommand{\betap}{\beta_+}
\newcommand{\betam}{\beta_-}
\newcommand{\fsix}{{\cal F}^{\perp}}

\begin{document}
\setcounter{tocdepth}{3}

%%%JHEP title
 
 \title{Constrained superfields on metastable anti-D3-branes}
 \author[a]{Lars Aalsma,}
 \author[a]{Jan Pieter van der Schaar,}
 \author[b]{Bert  Vercnocke}
 
\affiliation[a]{Institute for Theoretical Physics Amsterdam, Delta Institute for Theoretical Physics,\\
University of Amsterdam, Science Park 904, 1098 XH Amsterdam, The Netherlands}

 \affiliation[b]{Institute for Theoretical Physics, KU Keuven,\\
 Celestijnenlaan 200D, bus 2415, 3001 Leuven, Belgium}
 
\emailAdd{l.aalsma@uva.nl}
\emailAdd{j.p.vanderschaar@uva.nl}
\emailAdd{bert.vercnocke@kuleuven.be}
 
 \abstract{We study the effect of brane polarization on the supersymmetry transformations of probe anti-D3-branes at the tip of a Klebanov-Strassler throat geometry. As is well known, the probe branes can polarize into NS5-branes and decay to a supersymmetric state by brane-flux annihilation. The effective potential has a metastable minimum as long as the number of anti-D3-branes is small compared to the number of flux quanta. We study the reduced four-dimensional effective NS5-brane theory and show that in the metastable minimum supersymmetry is non-linearly realized to leading order, as expected for spontaneously broken supersymmetry. However, a strict decoupling limit of the higher order corrections in terms of a standard nilpotent superfield does not seem to exist. We comment on the possible implications of these results for more general low-energy effective descriptions of inflation or de Sitter vacua. 
 }
 \maketitle

 %%%%%%%%%%%%%%%%%%%%%%%%%%%%%

\newpage

\section{Introduction}
De Sitter vacua are at the heart of any cosmological model as both the early and late universe are well-approximated by a de Sitter phase. It is therefore of great importance to understand the construction of de Sitter vacua in string theory and supergravity. However, such constructions have proven to be a tremendous challenge. Kachru, Kallosh, Linde and Trivedi (KKLT) provided a generic mechanism of moduli stabilization in Anti-de Sitter and an uplift to de Sitter vacua in ten-dimensional string theory already in 2003 \cite{Kachru:2003aw} and by now, many different approaches for de Sitter compactifications have been uncovered. 

In contrast, the equivalent mechanism for de Sitter vacua in an effective four-dimensional $N=1$ supergravity theory was only developed recently, using constrained superfields \cite{Antoniadis:2014oya,Dudas:2015eha,Bergshoeff:2015tra,Hasegawa:2015bza,Cribiori:2016qif}. By imposing constraints on superfields it is not only possible to describe fields transforming non-linearly under the broken supersymmetry, but also to eliminate unwanted degrees of freedom. General prescriptions for constrained superfields from linearly transforming ones in a supergravity context were given in \cite{DallAgata:2016syy,Ferrara:2016een}. For some recent reviews of constrained superfields and their applications to cosmology, see \cite{Ferrara:2015cwa,Ferrara:2016ajl}.

Constrained superfields are often effective descriptions of the low-energy excitations. For example, in the context of four-dimensional spontaneous supersymmetry breaking the massless goldstino can be packaged in a chiral superfield that satisfies a nilpotency constraint. This constraint arises after the bosonic superpartner of the goldstino (the sgoldstino) becomes heavy enough to be integrated out \cite{Komargodski:2009rz,Kallosh:2015pho,Kallosh:2016hcm}. As argued in \cite{Komargodski:2009rz} this can be extended to multiple fields. In general, integrating out additional heavy degrees of freedom results in extra constraints which describe the universal low-energy dynamics of the theory, see also \cite{DallAgata:2016syy,Kallosh:2016hcm}.

It is of crucial importance to understand the embedding of constrained superfields in a putative UV-complete description. Can we indeed realize large mass splittings such that the constrained superfields correspond to a good approximation of the relevant low-energy physics? This question is especially important when considering cosmological inflation. As one typically accesses high energy scales during inflation it is necessary to ensure that the fields eliminated by the constraints have large enough masses to be integrated out. Otherwise, a constrained superfield description will be invalid.

An important condition for obtaining universal (UV insensitive) couplings to the goldstino, and standard constrained superfield descriptions, is that the masses of the heavy superpartners should be large compared to the supersymmetry breaking scale. If that condition is not fulfilled, the constraints are higher-order and depend on the masses of the heavy fields \cite{Dudas:2011kt}. This issue was recently reconsidered featuring global supersymmetry \cite{Ghilencea:2015aph} and supergravity \cite{Dudas:2016eej}. Those authors studied the emergence of the constraints by integrating out massive fields, instead of imposing the constraints by hand. In \cite{Dudas:2016eej} the corrections to an inflationary model with two superfields were analyzed. One superfield was used to describe spontaneous supersymmetry breaking and a second one contained the inflaton and its superpartner. The UV physics is described by a supergravity model with additional heavy superfields and supersymmetry is broken by an O'Raifeartaigh-mechanism. This particular UV model did not allow for an exact nilpotent superfield description, because the strict infinite mass limit of the sgoldstino that would decouple its fluctuations as in \cite{Kallosh:2015pho} does not exist. Instead, corrections due to the finite sgoldstino mass during inflation significantly limit the range of parameters for which an effective nilpotent description is available. It is not clear whether more generic UV models have similar restrictions on taking the large sgoldstino mass limit.  

In this paper we take a step back from inflation and study how universal the description of de Sitter vacua with a nilpotent superfield is, in the context of string theory. We build on the recent connection between constrained superfields in four-dimensional effective $N=1$ supergravity and string theory. The uplift term of the KKLT mechanism is generated by anti-D3 branes in a Giddings-Kachru-Polchinski (GKP) background \cite{Giddings:2001yu}. This uplift is an example of the generic string theory mechanism of supersymmetry breaking  by branes in backgrounds with fluxes.  If the anti-D3-brane indeed breaks supersymmetry spontaneously \cite{Kachru:2002gs,McGuirk:2012sb,Bertolini:2015hua} it should be possible to package a worldvolume fermion into a nilpotent superfield describing the goldstino. This expectation was confirmed explicitly by putting the anti-brane on top of an orientifold 3-plane in an $N=1$ flux background \cite{Kallosh:2014wsa,Bergshoeff:2015jxa,Vercnocke:2016fbt}.

The effective  description for the first constrained superfield models in the context of KKLT arises by explicitly putting the anti-D3-brane on top of the orientifold plane. To answer the question if a constrained superfield description of de Sitter vacua is still appropriate in a more general background, we remove the orientifold projection. One of us initiated this study with Kallosh and Wrase for a ten-dimensional flat background \cite{Kallosh:2016aep}: the non-linear transformations for all massless worldvolume fields (vector, scalars, fermion) can indeed be described by constrained multiplets. 

The full understanding of anti-D3-branes in flux backgrounds should introduce corrections to the description in the flat background of \cite{Kallosh:2016aep}. Anti-D3-branes at the bottom of a warped throat can polarize into NS5-branes under the influence of background flux \cite{Kachru:2002gs}. In this paper we show that one source of corrections is due to such polarization effects.\footnote{In recent years the literature has been divided on whether metastable anti-D3 probes are robust beyond probe level, for recent work see \cite{Bena:2014jaa,Michel:2014lva,Cohen-Maldonado:2015ssa,Bena:2016fqp} and references. We want to discuss the  appearance of non-linear supersymmetry and possible corrections first at probe level and do not discuss back-reaction in this paper.} We write down the supersymmetric version of the action for the polarized brane and consider small fluctuations around the metastable minimum from the four-dimensional point of view. This reveals that supersymmetry is indeed, to leading order in fluctuations, non-linearly realized at the minimum. The central question is at what scale the first leading corrections  to the standard four-dimensional non-linear description appear. We find that this scale is not set by the mass of the scalar fluctuations, but is instead smaller by a factor $p/M$, with $p$ the number of anti-branes and $M$ the flux number of the Klebanov-Strassler background. Interestingly, the strict limit that would decouple these corrections does not exist. That limit is equivalent to sending the dimensionless ratio $p/M$ to infinity, while anti-D3-branes only settle into a metastable state for sufficiently small $p/M$.

The rest of this paper is organized as follows.
We review the potential for polarized anti-D3 branes from the perspective of the NS5 worldvolume theory in section \ref{sec:bosKPV}, with special emphasis on the expected scale at which this description is valid. In section \ref{sec:fermKPV}, we construct the supersymmetric completion of the polarized NS5-brane action. We analyze the four-dimensional supersymmetry transformations in section \ref{sec:susytransf}. 
Finally, in section \ref{sec:conclusions} we comment on our findings and the relation to the use of anti-branes in de Sitter uplifts. Appendix \ref{app:spinors} contains a technical derivation of the fermionic terms in the action and the supersymmetry  transformations, based on the S-dual D5-brane action in a flux background of \cite{Martucci:2005rb}.

\section{The bosonic KPV potential}\label{sec:bosKPV}

\begin{figure}[ht!]
\centering
\includegraphics[width=.6\textwidth]{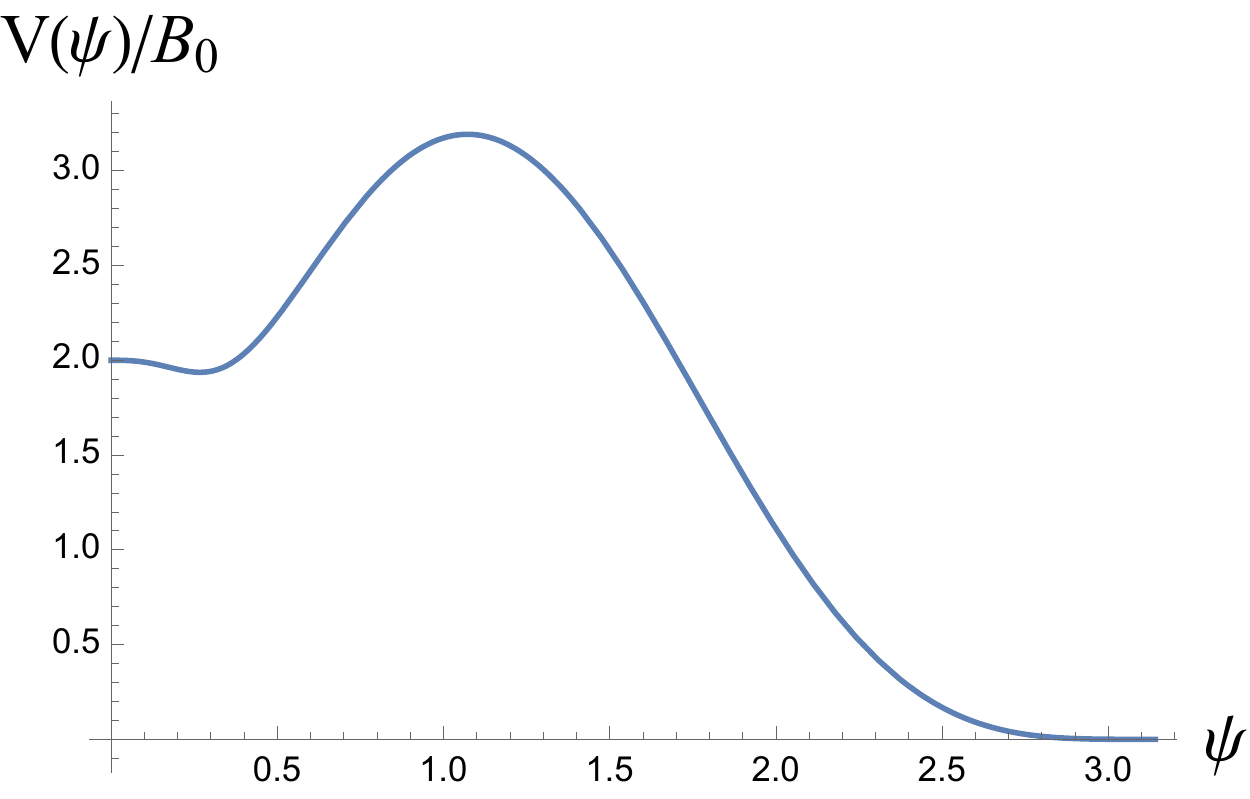}
\caption{The NS5-brane polarization potential for $p/M=0.03$ and $B_0\equiv e^{4A_0}p\,\mu_3/g_s$. At the north pole ($\psi=0$) there are p anti-D3-branes that polarize into an NS5-brane carrying charge $Q(\psi)$ at the metastable minimum ($\psi_{\rm min}$). This configuration can decay non-perturbatively to the supersymmetric minimum at the south pole ($\psi=\pi$), where the $p$ anti-D3-branes annihilated and the final configuration contains $(M-p)$ D3-branes.}
\label{fig:NS5pol}
\end{figure}

Let us start with a short review of some of the results of Kachru, Pearson and Verlinde (KPV) \cite{Kachru:2002gs}. KPV added $p$ anti-D3-branes to the warped deformed conifold geometry of Klebanov and Strassler \cite{Klebanov:2000hb}. The throat of this geometry is supported by $M$ units of flux through the A-cycle and $K$ units through the B-cycle.
\be
\frac{1}{4\pi^2}\int_A F_3=M \hspace{2em} \frac{1}{4\pi^2}\int_B H_3=-K
\ee
The Klebanov-Strassler geometry is an example of a GKP background \cite{Giddings:2001yu} that experiences a high degree of warping near the bottom of the throat in the six-dimensional geometry. Since probe anti-D3 branes in the Klebanov-Strassler background feel a net force towards the bottom of the throat we can describe the relevant physics by focusing on the region near the tip of the throat, with topology $R^4\times S^3$. The metric near the tip is \cite{Kachru:2002gs}
\be
ds^2=e^{2A_0}\eta_{\mu\nu} dx^\mu dx^\mu+ g_s M b_0^2 (d\psi^2+\sin^2\psi \,d \Omega_2^2) \, , \qquad b_0^2 \approx 0.93266\,.
\label{eq:metric}
\ee
with $e^{2A_0} = \varepsilon^{4/3}/g_s M$ the constant warp factor at the tip and $\varepsilon$ the deformation parameter of the deformed conifold.
Anti-branes carry opposite charge with respect to the supersymmetric background, breaking all supersymmetry. By brane polarization \cite{Myers:1999ps}, the anti-branes can blow up to form an NS5-brane wrapping an $S^2$ inside the $S^3$. Depending on the value of $p/M$ the NS5-brane either settles at a metastable minimum at a fixed radius of the $S^2$, or shrinks all the way to the opposite south pole of the $S^3$, brane-flux annihilation takes place and the final configuration becomes supersymmetric, see figure \ref{fig:NS5pol}.
Since the non-supersymmetric and supersymmetric states are continuously connected by moving the NS5-brane from the north pole to the south pole on the $S^3$, one expects the breaking of supersymmetry in the metastable vacuum to be spontaneous and supersymmetry to be realized non-linearly. We opt to describe the dynamics from the perspective of the effective NS5-brane worldvolume theory. The bosonic action describing the NS5 worldvolume theory is given by\footnote{The effective action on the NS5-brane is obtained by S-duality of the D5-brane DBI theory. Strictly speaking this description is therefore only valid for large $g_s$, but some (supersymmetric) properties and structures are expected to be invariant.}
\be
S_{\rm NS5}=\frac{\mu_5}{g_s^2}\int d^6\xi \sqrt{-\det G_\parallel}\sqrt{\det(G_\perp+2\pi g_s\mathcal{F})}+\mu_5\int B_6 \, .
\label{eq:NS5action}
\ee
We wrote the DBI term in terms of the metric components $G_\parallel$ spanned by the (anti-)D3 brane coordinates (Minkowski coordinates plus possibly motion in $\psi$) and $G_\perp$, spanned by the coordinates on the $S^2$. The form fields in the action are
\be
2\pi\mathcal{F}_2=2\pi F_2-C_2 \, , \qquad dB_6=-\frac{1}{g_s}dV_4\wedge F_3 \, .
\ee
The gauge field $A$ on the worldvolume has field strength $dA=F_2$ and gives the anti-D3 charge $p$ carried by the NS5-brane:
\be
 \int_{S^2} F_2 = 2 \pi p\,,
\ee
and $F_3 = dC_2$.

We are specifically interested in the effective dynamics in the angular direction $\psi$ on the $S^3$, which is transverse to the NS5-brane wrapped on an $S^2$ inside the $S^3$. The action becomes\footnote{Where we added a constant to the action such that the potential is zero at the supersymmetric minimum.}
\be
 S=  \frac{\mu_3}{g_s} e^{4A_0} \int d^4 \xi \left[\sqrt{P^2 + Q^2}\sqrt{1 + e^{-2A_0}g_{\psi\psi} \partial_\mu \psi \partial^\mu\psi} -  Q\right] \, ,
\ee
where we used $4\pi^2\mu_5=\mu_3$ and we introduced shorthand notation for the following integrals over the $S^2$:
\begin{eqnarray}
 Q(\psi) &\equiv -\frac 1{2\pi}\int_{S^2} {\cal F}_2 &= -p+\frac{M}{\pi}\left(\psi-\frac{1}{2}\sin(2\psi)\right)\,,\\
 P(\psi) &\equiv \frac 1 {4\pi^2 g_s}\int_{S^2}\sqrt {G_\perp}&= \frac{b_0^2M}{\pi}\sin^2\psi\,.
\end{eqnarray}
$Q$ describes the effective D3-charge at position $\psi$.

From the action one can find the potential (the Hamiltonian at zero momentum):
\begin{align}
  V(\psi)&=\frac{\mu_3}{g_s}e^{4A_0}  \left(\sqrt{Q^2+P^2}-Q\right) \nonumber \\
 &=\frac{\mu_3}{g_s}e^{4A_0} \sqrt{Q^2+P^2} \left(1 + \cos (\alpha)\right) \label{eq:VDBI} \,,
\end{align}
where for later convenience we introduced the position-dependent angle $\alpha(\psi)$, which takes values $\alpha = \pm 1$ at the poles of the $S^3$:
\begin{equation}
\cos(\alpha(\psi)) \equiv -\frac{Q(\psi)}{\sqrt{P(\psi)^2+Q(\psi)^2}}\, , \qquad \sin(\alpha(\psi)) \equiv  -\frac{P(\psi)}{\sqrt{P(\psi)^2+Q(\psi)^2}}  \, .\label{eq:alpha}
\end{equation}
We plot the potential in figure \ref{fig:NS5pol}. It has a metastable minimum for relatively small values of $\psi$.

By expanding for small values of $\psi$ up to fourth order we find
\begin{equation}
V(\psi)\simeq \frac{p\mu_3}{g_s}e^{4A_0} \left(2-\frac{4M}{3\pi p}\psi^3+\frac{b_0^4M^2}{2\pi^2p^2}\psi^4\right) \, .
\label{eq:Vsmall}
\end{equation}
Around the north pole of the $S^3$ ($\psi=0$) the scalar field fluctuations are massless and we note that this state is unstable. For $p/M< 0.08$ there is a metastable minimum at
\be
\psi_{\rm min}=\frac{2\pi p}{b_0^4 M} \, .
\ee
In a moment we will expand the action in fluctuations $\delta\psi$ around the minimum $\psi_{\rm min}$. A general expansion of an arbitrary potential in fluctuations $\delta\phi$ up to cubic order can be written as 
\be
V(\phi_{\rm min}+\delta\phi) = V(\phi_{\rm min}) + \frac{1}{2}m_\phi^2(\delta\phi)^2 + \frac{\lambda_3}{3}(\delta\phi)^3 \, .
\ee
Using the standard normalization in four dimensions, a scalar field $\phi$ and the cubic coupling $\lambda_3$ have mass dimension 1. For fluctuations $\delta \phi < m_\phi^2 / \lambda_3$ the quadratic term is a good approximation of the relevant physics, but for larger fluctuations the cubic term dominates, signaling a breakdown of the quadratic approximation. For natural couplings ($\lambda_3 \sim m_\phi \sim \Lambda$), where $\Lambda$ is some (high-energy) cut-off scale, this would just restrict the field fluctuations to values below $m_\phi \sim \Lambda$, but for `unnaturally' large cubic couplings the quadratic approximation would only be valid for fluctuations significantly smaller than the mass scale $m_\phi$. 

The second situation is exactly what we observe in the Klebanov-Strassler throat. Expanding \eqref{eq:Vsmall} up to cubic order around $\psi_{\rm min}$ we obtain
\be
V(\psi_{\rm min}+\delta\psi) =\frac{p\mu_3}{g_s}e^{4A_0} \left( 2 -\frac{2}{3 }b_0^{-4}\psi_{\rm min}^2  + 4b_0^{-4}(\delta \psi)^2 +\frac{16}{3}b_0^{-4} \frac{(\delta \psi)^3}{\psi_{\rm min}} \right) \, . \label{eq:VDBIfluc}
\ee
From this expression, we see that $\lambda_3/m_\psi = 4\sqrt{2} / (b_0^2\psi_{\rm min})$. The cubic coupling is $\mathcal{O}(M/p)$ larger than the quadratic coupling. In the remainder of this article, we are  interested in the quadratic approximation. We are then forced to restrict to fluctuations that are not only small compared to the dimensionless mass parameter $m_\psi = 2\sqrt{2}/b_0^2$, which is of order one, but small compared to a dimensionless parameter set by the field value in the metastable minimum:
\be
\delta \psi \ll \psi_{\rm min} \sim p/M\,.
\ee
The importance of this basic observation will become clear when we discuss the corrected supersymmetry transformations in the metastable vacuum. 

In the rest of this paper, we continue with the discussion of the fermions on the NS5 worldvolume. We will be concerned with the leading behaviour at a fixed but small value of $p/M$. Then we consider the small $\psi$ expansion, and discuss small field fluctuations around a fixed background position for small $\psi$. We consider up to  quadratic order in the scale $\psi_{\min} \sim p/M$ and consider quadratic fluctuations in fields only.

From the action it is straightforward to obtain the potential for the canonically normalized field. For small fluctuations around a minimum at $\psi\ll1$, the kinetic term gets a constant prefactor as $\sqrt {P^2 + Q^2}  = p + O(\psi^3)$. We then find
\be
S_{\rm NS5}= e^{4A_0}\frac{p \mu_3}{g_s} \int d ^4 \xi  \left(\frac12 \partial_\mu \psi \partial^\mu\psi + \tilde{V}(\psi) \right) \, , \label{eq:SDBI}
\ee
with the potential
\be
\tilde{V}(\psi) = p^{-1}  \left(\sqrt{Q(\psi)^2+P(\psi)^2}-Q(\psi)\right) \, . \label{eq:VDBIb} 
\ee
We will arrange the kinetic terms of the fermions to have the same constant prefactor (for small fluctuations at least), such that we can consistently compare mass scales.

\section{The fermionic KPV potential}\label{sec:fermKPV}

Since we expect the metastable minimum to break supersymmetry spontaneously, there should exist an associated massless goldstino. For a single anti-D3-brane on top of an orientifold plane, the goldstino was identified as the 4d fermion on the worldvolume of the anti-brane, which is a singlet under the $SU(3)$ holonomy of the 6d internal space \cite{Kallosh:2014wsa,Bergshoeff:2015jxa}. Removing the orientifold plane, we now want to revisit the situation for the polarized NS5-brane. Based on the physical picture of the previous section, we expect the effective 4d worldvolume description to reduce to the known results for $p$ anti-D3-branes at the north pole and $M-p$ D3-branes at the south pole, both probing the GKP background.

\subsection{The fermionic action up to second order}

Just as for the bosonic action, we formally obtain the fermionic NS5-brane worldvolume action from S-duality of a D5-brane. The action up to quadratic order in fermions is given by \cite{Martucci:2005rb} (notice that we have a background with a constant dilaton) 
\be\label{eq:action}
S_{\rm NS5}=\frac{1}{2}\frac{\mu_5}{g_s^2}\int d^6\xi \sqrt{-\det(g+2\pi g_s\mathcal{F})}\bar{\theta}(1-\Gamma_{\rm NS5})\left[(\tilde M^{-1})^{\alpha\beta} \hat \Gamma_\beta D_\alpha-\Delta\right]\theta \, ,
\ee
where
\begin{align}
\tilde M_{\alpha\beta} &=g_{\alpha\beta}+2\pi g_s \sigma_3 \mathcal{F}_{\alpha\beta} \,, \nonumber \\
D_\alpha&=\nabla_\alpha+ W_\alpha \,, \nonumber \\
W_\alpha&=\frac{1}{8}\left(-F_{\alpha np}\Gamma^{np}\sigma_3+\frac{1}{3!}g_s^{-1} H_{mnp}\Gamma^{mnp} \hat \Gamma_{\alpha}\sigma_1\right) \,, \nonumber\\
\Delta&=\frac{1}{24}\left(-F_{mnp}\,\sigma_3-g_s^{-1} H_{mnp}\,\sigma_1\right)\Gamma^{mnp} \, .
\label{eq:D5actionterms}
\end{align}
We only included terms in the action that are non-zero at the tip of the throat, because we are not interested in dynamics taking us away from the tip (we dropped terms with five-form and one-form field strengths). The indices $m,n$ are ten-dimensional curved indices, $\alpha,\beta$ indicate worldvolume indices. To avoid confusion with the equations below, we wrote the pullbacks of gamma matrices on the worldvolume with hats: $ \hat \Gamma_\alpha=\Gamma_{\underline m}e_m{}^{\underline m}\partial_\alpha x^m$ and we underline tangent space indices ($\underline{m},\underline{n}\ldots$). The fermion $\theta$ is a doublet of Majorana-Weyl spinors with positive chirality.

We now use the specific embedding of the NS5-brane of the previous section and use the leg structure of the three-forms to simplify the expressions. The $F_3$ flux is fully along the $S^3$ spanned by $(\theta,\phi,\psi)$ while $H_3$ is orthogonal to $F_3$ in the internal space. This means we can drop $H_3$ terms with legs along the worldvolume of the NS5-brane. Also we will drop the terms with $\partial_\alpha \psi$ coming from the pullbacks of gamma matrices, as those do not contribute to the mass matrix. We only highlight the main points of the calculation here. For more general expressions and more detailed information, see appendix \ref{app:spinors}.

The combination in right brackets of \eqref{eq:action} gives:
\begin{align}
\label{eq:step1}
(\tilde M^{-1})^{\alpha\beta}\Gamma_\beta D_\alpha&-\Delta\cr
&= (\tilde M^{-1})^{\alpha\beta}\Gamma_\beta \nabla_\alpha- \frac 1 {24}\left( -\cos (2\alpha) F_{mnp} \sigma_3 +  (1 + \sin^2 \alpha) g_s^{-1} H_{mnp}\sigma_1\right)\Gamma^{mnp}\cr
&+ \cos^2(\alpha)(( {2 \pi g_s \sigma_3\cal F})^{-1})^{\alpha\beta}\left(-\frac 1 {8\cdot 3!}g_s^{-1} H^{npq} \sigma_1\Gamma_{\alpha\beta npq}-\frac 14 F_{\alpha \beta q}\sigma_3 \Gamma^{q}
\right)\,,
\end{align}
with the position-dependent angle $\alpha$ defined in \eqref{eq:alpha}.

It is important for our calculations to note that $\Gamma_{\rm NS5}$ is off-diagonal. As explained in appendix \ref{app:spinors}, at the tip of the deformed conifold, this projector takes on a fairly simple form:
\be
\Gamma_{\rm NS5} = -
\begin{pmatrix}
0 & \betam \\
\betap & 0
\end{pmatrix}
\,,
\qquad
\beta_\pm = \Gamma_{\underline{0123}} (\pm\cos(\alpha)  -  \Gamma_{\underline{45}}\sin(\alpha)) \,.
\ee
We still need to gauge fix the kappa-symmetry on the brane. We do this by taking the gauge fixing condition on the doublet $\theta=(\theta_1,\theta_2)$
\be
\sigma_3 \theta = -\theta \quad \Rightarrow \quad \theta_1 = 0 \, .
\ee
Now we can express the action in terms of the spinor $\theta_2$ only. This gauge fixing condition is convenient due to its simplicity, but it is not suitable when one also wants to perform an orientifold projection. The calculation for the mass matrix can also be done in a gauge where we set $(1+\Gamma_{\rm NS5})\theta=0$, compatible with an orientifold. We show in the appendix that this choice of gauge does not change the mass matrix.

We introduce the notation for the remaining spinor components
\be
\lambda \equiv \theta_2\,.
\ee
Taking care of the off-diagonal matrix $\Gamma_{\rm NS5}$ and using that for a 10d Majorana-Weyl spinor $\lambda$ the only fermion bilinears that are non-zero have three or seven gamma matrices, we find the result
\be
\label{eq:action_mass_matrix}
S_{\rm NS5}=\frac{p\mu_3}{g_s}e^{4A_0}\int d^4 \xi\, \int \frac{d\Omega}{4\pi} \, \bar\lambda [({\tilde M}^{-1})^{\mu\nu}\Gamma_\nu\nabla_\mu + {\cal M}]\lambda \, ,
\ee
with $d \Omega$ the volume element on the unit two-sphere.

The only terms that contribute to the mass matrix $\cal M$ are
\be
\label{eq:fermion_mass_matrix}	
{\cal M} = \frac 1 {24}\left( \cos (2\alpha) F_{mnp} - g_s^{-1}\cos (\alpha)  H_{mnp}\Gamma_{\underline{0123}} \right)\Gamma^{mnp} \,.
\ee
This is the mass matrix on the six-dimensional world volume. The reduction to four dimensions could also pick up extra mass terms coming from the reduction of the kinetic term \cite{Dasgupta:2016prs}. To determine if these extra mass terms still allow for a massless fermion, we have to make sure the internal piece of the modified Dirac operator together with the mass matrix $[({\tilde M}^{-1})^{\alpha\beta}\Gamma_\alpha\nabla_\beta + {\cal M}]$ has a zero mode.

In the remainder of this section we show that this is indeed the case and the lowest Kaluza-Klein modes reveal the existence of a massless fermionic mode, which we will identify as the massless goldstino.

\subsection{Reduction to four dimensions} \label{subsub:4dreduction}
In the previous section we obtained the action for the worldvolume fermions from the six-dimensional point of view. We now discuss the four-dimensional interpretation. When we perform the reduction to four dimensions, we will write $\lambda$ in terms of four fermions: a singlet $\lambda^0$ and a triplet $\lambda^i$ under the $SU(3)$ holonomy of the six-dimensional transverse internal space. This decomposition can for instance be found in \cite{Bergshoeff:2015jxa}.

Let us first focus on the reduction of the mass matrix. We observe that, up to angles that parameterize the position of the NS5 on the $S^3$, it is completely determined by the flux of the background, which can be written in terms of the complexified three-form 
\be
 G_3 = F_3 -i  g_s^{-1} H_3\,.
\ee
Supersymmetry of the Klebanov-Strassler background dictates that $\star_6 H_3=-g_sF_3$ or equivalently that the complex three-form is imaginary self-dual (ISD) $G_3 = i \star_6 G_3$ \cite{Kachru:2002gs}.
This immediately implies that the only relevant structure for the fermionic mass matrix we have to reduce to four dimensions is the real part of the complex three-form:
\be
{\cal M} = \frac 1 {48}\left( \cos (2\alpha)+\cos(\alpha)\right )(G_3 + \bar G_3)_{mnp}\Gamma^{mnp}\label{eq:metastablemass0} 
\ee
Up to the coordinate-dependent prefactor $\left( \cos (2\alpha)+\cos(\alpha)\right )$, this is the known mass term for anti-D3 branes in a supersymmetric background with fluxes that carry only D3-brane charges, as reviewed in \cite{Bergshoeff:2015jxa}. The general discussion of our mass terms also carries through directly as in \cite{Bergshoeff:2015jxa}. The background three-form is (2,1) and primitive, and therefore we find that the only non-zero contributions to the mass matrix come from the triplet:
\be
 \bar \lambda {\cal M} \lambda = m_{ij} \bar \lambda^i_+ \lambda^j_+ + \bar m_{\bar \imath \bar \jmath} \bar \lambda^{\bar \imath}_- \lambda^{\bar \jmath}_-\,,
\ee
where the $m_{ij}$ are linear in the components of the background flux and  $\pm$ subscripts denote 4d Weyl spinors $\lambda_\pm = \frac 12 (1+ i \Gamma_{\underline{0123}})\lambda$. We thus find that the mass matrix only leaves $\lambda^0$ massless, similar to a single anti-D3-brane that does not polarize \cite{Bergshoeff:2015jxa}.

The kinetic term of the fermions still contains a `modified Dirac operator'
\be
({\tilde M}^{-1})^{\alpha\beta} \Gamma_\alpha\nabla_\beta \lambda  =((g + 2 \pi g_s\sigma_3 {\cal F})^{-1})^{\alpha\beta} \Gamma_\alpha\nabla_\beta \lambda  
\ee
that could contribute to the mass matrix in four dimensions. We can ask whether there is a fermion that remains massless and signals the spontaneous breaking of supersymmetry. The flux is crucial. If we would reduce the Dirac operator on an $S^2$ without worldvolume flux ${\cal F}$, this would leave no fermion massless, as the 2-sphere admits no covariantly constant spinors. However, we have a non-zero worldvolume flux ${\cal F}$ on the $S^2$ that induces the (anti-)D3 brane charge. This allows for the possibility that the gauge field twists the Dirac operator on the $S^2$ such that the modified Dirac operator can have a zero mode on the 2-sphere, along the lines of \cite{Andrews:2005cv}. If that zero mode agrees with the $\lambda^0$ direction, we can identify $\lambda^0$ as the four-dimensional goldstino, as was suggested in \cite{McGuirk:2012sb}. 

Instead of explicitly solving $({\tilde M}^{-1})^{\alpha\beta} \Gamma_\alpha\nabla_\beta \lambda  = 0$, we will opt to describe this massless mode from the dual perspective of the non-abelian gauge theory on the anti-D3 branes. From this point of view the situation is more transparent because a reduction to four dimensions is not needed.
In the non-abelian theory all fields become matrix-valued. The transverse scalars $\phi^i$ have a potential that describes the brane polarization. One finds that the local minimum of this potential occurs when the scalars take an irreducible representation of $SU(2)$ \cite{Kachru:2002gs}, which agrees with the metastable minimum of the wrapped NS5-brane.  In the non-abelian theory, the analogue of the abelian 2-sphere with coordinates $\theta,\phi$ is a non-commutative fuzzy 2-sphere. 

The non-abelian theory is studied in detail in \cite{McGuirk:2012sb}, with a decomposition of the 10d worldvolume fermion $\lambda$ (which is promoted to a matrix) to the 4d fields $\lambda^0$ (`gaugino') and $\lambda^i$ (`modulini'), analogous to the abelian theory. For supersymmetry preserving ISD $G_3$-flux, the gaugino mass terms vanish and only the modulini are massive, in agreement with our NS5 mass matrix ${\cal M}$. An additional mass contribution might come from Yukawa couplings between $\phi^i$, $\lambda^0$ and $\lambda^i$:
\be
[\phi^{\bar \jmath} , \lambda^i] \lambda^0 + h.c.
\ee
The scalars $\phi^i$ have a vacuum expectation value in the metastable minimum such that the Yukawa coupling can be viewed as an off-diagonal contribution to the mass matrix. To find the massless goldstino, we expand the fermions in terms of eigenfunctions on the fuzzy sphere (the non-commutative analogue of spherical harmonics \cite{Iso:2001mg}). One finds that the lowest ($\ell=0$) mode of $\lambda^i$ commutes and that its corresponding Yukawa coupling vanishes, leaving $\lambda^0$ massless. Higher $(\ell>0)$ modes correspond to a Kaluza-Klein tower \cite{Andrews:2005cv} and can be ignored when the radius of the fuzzy sphere is sufficiently small. Clearly, ignoring $\ell>0$ modes we are left with an abelian truncation of the non-abelian fermionic action where we can identify $\lambda^0$ as a goldstino. This verifies the idea that in this setup spontaneous supersymmetry breaking should come with a massless fermion.

\subsection{Mass matrix in four dimensions}

To facilitate comparison with similar treatments in the literature, we will now explicitly compute the mass matrix $\cal M$ at the three relevant positions: the two poles of the $S^3$ and the metastable minimum at $\psi = \psi_{\rm min}$.  As mentioned before, we can rewrite the mass matrix in terms of the complexified three-form $G_3$. The general form of the mass matrix then becomes
\be
{\cal M} = \frac 1 {48}\left( \cos (2\alpha)+\cos(\alpha)\right )(G_3 + \bar G_3)_{mnp}\Gamma^{mnp}\label{eq:metastablemass1}
\ee
We now give the four-dimensional reduction and discuss the fermionic mass matrix on the positions of interest.

\subsubsection{Mass matrix at the poles}

At the North pole, $\psi=0$, we have  $p$ anti-D3 branes with $\cos(\alpha) =+1$. At the South pole, we have $M-p$ D3-branes at the supersymmetric minimum and $\cos(\alpha) = -1$.

The mass matrix ${\cal M}$ becomes
\begin{align}
{\cal M}(\psi=0) &= \frac{1}{24}\left(G_3+\bar{G}_3\right)_{mnp}\Gamma^{mnp}\,,\\
{\cal M} (\psi=\pi)&=0 \, .
\end{align}
These match  earlier results for anti-D3 branes or D3 branes on GKP backgrounds derived in \cite{Grana:2002tu,Bergshoeff:2015jxa} (note that we are working in an S-dual frame compared to those references, so one should take $G_3 \to -i g_s^{-1} G_3$ for comparison to those references.)

\subsubsection{Mass matrix at the metastable minimum}

To obtain the mass matrix in the metastable minimum, we expand $\cos(\alpha)$ to lowest non-trivial order and we evaluate this expression at the minimum:
\be
{\cal M} = \frac1{24}\left(1-\frac54\alpha^2 \right)\left(G_3+\bar{G}_3\right)_{mnp}\Gamma^{mnp}\, .
\ee
From the small $\psi$ expansion of $\alpha$ we have $\alpha = \frac{b_0^2}{\pi} \frac{M}p \psi^2 + O(\psi)^4$. To lowest order in $p/M$ we find
\be
\alpha(\psi_{\rm min})=\frac{4 \pi}{b_0^{6}}\frac pM.
\ee
In terms of $G_3$ flux we have the mass matrix
\be
{\cal M}=\left(\frac{1}{24}-\frac56\frac{\pi^2}{b_0^{12}}\frac {p^2}{M^2}\right)\left(G_3+\bar{G}_3\right)_{mnp}\Gamma^{mnp} \, .
\label{eq:metastablemass}
\ee

\section{Supersymmetry transformations}\label{sec:susytransf}

In the previous section we argued that in the metastable minimum supersymmetry is spontaneously broken by identifying the corresponding massless goldstino. This also suggests that the effective low-energy dynamics can be described in terms of a nilpotent superfield \cite{Komargodski:2009rz}. In this section we analyze the supersymmetry transformations to verify this picture and identify the leading corrections.  

To begin we need the expressions for the supersymmetry transformations in non-trivial flux backgrounds, which can be found in short in appendix \ref{app:spinors}, adapted from \cite{Martucci:2005rb}. Supersymmetry of the background requires that
\be
(1+i\sigma_2\Gamma_{\underline{0123}})\,\epsilon = 0 \quad \Leftrightarrow \quad  \epsilon_2 =\Gamma_{\underline{0123}}\epsilon_1  \, .
\ee
With a slight abuse of notation, we will write the 32-component Majorana-Weyl spinor again as $\epsilon \equiv -2\epsilon_2$.
We have the following supersymmetry transformations:
\begin{align}
\delta_\epsilon \lambda &= -\frac 12 [\unity - \beta] \epsilon + {\cal O}(\lambda)^2\,,\\
\delta_\epsilon \psi &= \frac 12 \bar \lambda\Gamma^\psi [\unity  + \beta]\epsilon+ \xi^\mu \partial_\mu \psi + {\cal O}(\lambda)^3\,, \\
\delta_\epsilon A_\mu &= -\frac 12 \bar \lambda ( \Gamma_\mu + \Gamma_\psi \partial_\mu \psi)[\unity  + \beta]  \epsilon + \frac 12 C_{\mu m} \bar \lambda \Gamma^m [\unity  + \beta]  \epsilon+ \xi^\nu F_{\nu\mu}+ {\cal O}(\lambda)^3\,,
\end{align}
with $\xi^\mu = -\frac 12 \bar \lambda \Gamma^\mu (\unity + \beta) \epsilon$ and the operator $\beta$ defined as $\beta = \Gamma_{\underline{0123}} \betap$, see eq.\ \eqref{eq:betas}:
\begin{align}
  \beta &= - \Big{(}\cos (\alpha) - \Gamma_{\underline{45}} \sin (\alpha)\Big{)}\Big{(} 1  + \frac 12 \ffourtwo _{\mu\nu}\Gamma^{\mu\nu}+\partial^{\mu} \psi \Gamma_{\psi\mu}- \frac 12 g_{\psi\psi}\partial_\mu\psi\partial^\mu \psi
 - \frac 14 \ffourtwo_{\mu\nu} \ffourtwo^{\mu\nu}\cr
 &\quad\, + \frac 14 \ffourtwo_{\mu\nu} (\star_4 \ffourtwo)^{\mu\nu} \Gamma_{\underline{0123}}+\frac 12 \partial_{\rho}\psi \ffourtwo _{\mu\nu}\Gamma^{\psi\rho\mu\nu} + \ldots\Big{)}
\end{align}
We do not write fermion terms in $\beta$, as those result in transformations that take use beyond the quadratic fermion order in the action.

More details on these transformations can be found in appendix \ref{app:spinors}. From here, we can already see the general form of the transformations around the poles, since
\begin{align}
\text{at } \psi &= 0: \quad \beta=-\unity+\ldots\,, \nonumber  \\
\text{at } \psi &= \pi: \quad \beta = +\unity+\ldots \, ,
\end{align}
where the ellipses denote terms with field fluctuations. So around $\psi=0$ we find non-linear transformations and at $\psi=\pi$ linear ones.

To obtain four-dimensional supersymmetry transformations, in the end we  always decompose the spinor into the singlet $\lambda^0$ and the triplet $\lambda^i$ under the $SU(3)$ holonomy. Moreover we can focus on just one of the triplet fermions, say $i=3$, due to the arbitrary orientation of the $S^2$ inside the transverse $S^3$, corresponding to the superpartner of the scalar $\psi$ at the south pole where supersymmetry is restored. The other directions come along for the ride and we will ignore them throughout.
We are also interested in the supersymmetry transformations with parameter $\epsilon^0$, the SU(3)  singlet component of the 32-component Majorana-Weyl spinor $\epsilon$, as this is the supersymmetry preserved by the background.

With all the relevant information in place, we present a summary of the four-dimensional fermionic, scalar and gauge field supersymmetry transformations at the different locations of interest: both poles and most importantly the metastable minimum.

\subsection{At the south pole}

Let us first analyze the south pole $\psi = \pi$, where the D3-branes do not break the background $N=1$ supersymmetry. 
We obtain to leading order in fluctuations the expression for $\beta$:
\be
\beta =  \unity + \frac 12 \ffourtwo _{\mu\nu}\Gamma^{\mu\nu}+\partial_{\mu} \psi \Gamma_{\psi}\Gamma^\mu\,,
\ee
and the reduction of the supersymmetry transformations to four dimensions gives
\begin{align}
\delta_\epsilon \lambda^0 &= \frac14 \gamma^{\mu\nu} F_{\mu\nu}\epsilon^0 \,,\\
\delta_\epsilon \lambda^3 &= \frac{1}{\sqrt{2}}\gamma^{\mu}\partial_\mu\tilde\psi\epsilon^0 \,,\\
\delta_\epsilon \tilde\psi &= \frac{1}{\sqrt{2}}\bar{\lambda}^3\epsilon^0 \,,\\
\delta_\epsilon A_\mu &= -\frac12\bar{\lambda}^0\gamma_\mu\epsilon^0 \,,
\end{align}
where we redefined the scalar as follows.
\be
 {\tilde \psi} =  -e^{\underline \psi}_\psi\, \psi = -(g_s M b_0^2)^{1/2} \psi \,,
\ee
and rescaled spinors as $\lambda\to \frac{1}{\sqrt{2}}\lambda$, $\epsilon\to \frac{1}{\sqrt{2}}\epsilon$. We conclude that, as expected, at $\psi=\pi$ a linearly realized $N=1$ supersymmetry exists under which $(\lambda^0,A_\mu)$ form a vector multiplet and $(\lambda^3,\psi)$ correspond to a chiral multiplet. If we would have included the other two directions on the $S^2$ that we now have ignored, they would form two additional chiral multiplets. Those correspond to the decomposition of the spectrum of the $N=4$ SYM multiplet on the D3-brane transforming under the $N=1$ of the background.

\subsection{At the north pole}

At the (unstable) north pole we expect the effective description to formally reduce to the results for a supersymmetry breaking anti-D3-brane in a GKP background. We will write the transformations to at most quadratic order in field fluctuations. Since $\sin(\alpha) = {\cal O}(\psi^2), \cos(\alpha) = 1 + {\cal O}(\psi^4)$, we set $\cos \alpha=1$, as the subleading terms will come in at higher order in the supersymmetry transformations. Then we indeed reproduce to quadratic order the results of \cite{Kallosh:2016aep}.

We will expand the supersymmetry transformations up to the first non-trivial order in the fields. Then we only have to expand the operator $\beta$ to first order:
\be
\beta = -\unity -  \frac 12 \ffourtwo _{\mu\nu}\Gamma^{\mu\nu}-\partial^{\mu} \psi \Gamma_{\psi\mu}+ \ldots\,.\label{eq:beta_psi0}
\ee
The supersymmetry transformations around $\psi = 0$ are
\begin{align}
\delta_\epsilon \lambda &=-\epsilon -  
\frac 14 \ffourtwo _{\mu\nu}\Gamma^{\mu\nu}\epsilon-\frac 12 (\partial_{\mu} \psi) \Gamma_{\psi}\Gamma^{\mu}\epsilon + {\cal O}(\phi^2)\,, \nonumber\\
\delta_\epsilon \psi &= - \frac 12 (\bar \lambda  \Gamma^\mu \epsilon )\partial_\mu \psi -\frac14( \bar \lambda \Gamma^{\psi\mu\nu} \epsilon) F_{\mu\nu} + {\cal O}(\phi^3)\,, \nonumber\\
\delta_\epsilon A_\mu &= -\frac 12 (\bar \lambda \Gamma^\rho \epsilon) F_{\rho\mu} - \frac 12 (\bar \lambda \Gamma_\psi \epsilon)\partial_\mu \psi  + \frac 14 (\bar \lambda \Gamma_{\mu\rho\sigma} \epsilon ) F^{\rho \sigma}+ \frac 12 (\bar \lambda \Gamma_{\psi \rho \mu}\epsilon )\partial^\rho \psi + {\cal O}(\phi^3)\,,
\end{align}
with $\phi$ the collection of all fields $\phi = \{\psi,\lambda,A_\mu\}$.
We recognize the first terms as the standard non-linear transformations. By requiring the fields to transform non-linearly under the supersymmetry we can perform appropriate field redefinitions of the spinors, scalar and gauge field, that fix the transformations uniquely:
\begin{align}
\tilde \lambda &=- \lambda  + \frac 14 \ffourtwo _{\mu\nu}\Gamma^{\mu\nu}\lambda + \frac 12( \partial_{\mu} \psi )\Gamma_{\psi}\Gamma^{\mu}\lambda + {\cal O}(\phi^3)\nonumber \,,\\
\tilde \psi &= \psi  -\frac18( \bar \lambda \Gamma^{\psi\mu\nu} \lambda) F_{\mu\nu} + {\cal O}(\phi^4) \nonumber \,,\\
\tilde A_\mu&= A_\mu  - \frac 14 (\bar \lambda \Gamma_\psi \lambda)\partial_\mu \psi + \frac 18 (\bar \lambda \Gamma_{\mu\rho\sigma} \lambda ) F^{\rho \sigma} + \frac 14 (\bar \lambda \Gamma_{\psi \rho \mu}\lambda )\partial^\rho \psi  + {\cal O}(\phi^4)\,,\label{eq:northpole_fieldredef}
\end{align}
and we have the standard-looking transformations
\begin{align}
\delta_\epsilon \tilde\lambda &=\epsilon + {\cal O}(\phi^2)\,,\nonumber\\
\delta_\epsilon \tilde\psi &= \frac 12 (\bar{\tilde\lambda}  \Gamma^\mu \epsilon )\partial_\mu \tilde\psi + {\cal O}(\phi^3)\,,\nonumber\\
\delta_\epsilon \tilde A_\mu &= \frac 12 (\bar{\tilde\lambda} \Gamma^\rho \epsilon) \tilde F_{\rho\mu} + {\cal O}(\phi^3)\,.\label{eq:northpole_nonlinear}
\end{align}
With  an additional rescaling of the spinors $\tilde \lambda \to\sqrt 2 \tilde \lambda, \epsilon \to \sqrt{2} \epsilon$, we then find the following supersymmetry transformations in terms of the appropriate four-dimensional fields around $\psi=0$
\begin{align}
\delta_\epsilon \tilde\lambda^0 &= \epsilon^0 + {\cal O}(\phi^2) \\
\delta_\epsilon \tilde\lambda^3 &= 0 + {\cal O}(\phi^2) \\
\delta_\epsilon \tilde\psi &= (\bar{\tilde\lambda}^0\gamma^\mu \epsilon^0)\partial_\mu \tilde\psi + {\cal O}(\phi^3) \\
\delta_\epsilon  \tilde A_\mu &= (\bar{\tilde\lambda}^0\gamma^\nu \epsilon^0) \tilde F_{\nu\mu} + {\cal O}(\phi^3) \, .
\end{align}
We conclude that indeed, as anticipated by the physical interpretation in terms of brane-flux decay, this seems to describe an exact non-linear realization of (broken) supersymmetry when adding anti-D3-branes to the GKP background and ignoring the (higher order) dynamics describing the polarization in the transverse $S^3$ directions. This matches the results for anti-D3 branes in supersymmetric backgrounds of \cite{Kallosh:2014wsa,Bergshoeff:2015jxa,Vercnocke:2016fbt,Kallosh:2016aep}. Note that this (direct) expansion of the theory around the north pole is only a \emph{formal} result: since the scalar field $\psi$ sits at the maximum of its potential, this is an expansion around an unstable configuration.

\subsection{At the metastable minimum}

Now let us include the polarization dynamics and determine the transformations at the true metastable minimum $\psi_{\rm min}$, which should include corrections due to the dynamics on the $S^3$. We first expand in $\psi$ and then in the fluctuations around the metastable minimum. The expansion for $\alpha$ around the metastable minimum is then
\be
\alpha(\psi_{\rm min}+\delta\psi) = \frac{4\pi}{b_0^6}\frac{p}{M}+\frac{4}{b_0^2}\delta\psi+\frac{b_0^2}{\pi}\frac{M}{p}\delta\psi^2  + \ldots
\label{eq:alphaexp}
\ee
The leading corrections in the expansions of $\psi$-fluctuations and powers of $p/M$ are then captured by expanding $\beta$ in powers of $\alpha$:
\be
  \beta = \Big{(}1 - \alpha \Gamma_{\underline{45}} - \frac 12 \alpha^2  + \ldots\Big{)} \beta|_{\psi = 0}\,,
\ee
where $\beta|_{\psi = 0}$ is given by \eqref{eq:beta_psi0}.

We find that after the field redefinition \eqref{eq:northpole_fieldredef} and the spinor rescalings the transformations \eqref{eq:northpole_nonlinear} are corrected by the $\alpha$-expansion (or equivalently $\psi$-expansion):
\begin{align}
\delta_\epsilon \tilde \lambda &= \delta_\epsilon \tilde \lambda|_{\psi = 0} - \frac 12  \alpha \Gamma_{\underline{45}} \epsilon - \frac 14 \alpha^2 \epsilon+ \ldots \,,\cr
\delta_\epsilon \tilde \psi &=  \delta_\epsilon \tilde\psi|_{\psi = 0} -  \alpha \bar{\tilde\lambda}\Gamma^{\underline{45}\psi}\epsilon - \frac 12 \alpha^2\bar{\tilde\lambda}\Gamma^\psi \epsilon + \ldots \,,\cr
 \delta_\epsilon \tilde A_\mu  &= \delta_\epsilon \tilde A_\mu |_{\psi = 0} + \alpha \bar{\tilde\lambda}\Gamma_{\underline{45}}\Gamma_{\mu}\epsilon + \frac 12 \alpha^2\bar{\tilde \lambda}\Gamma_\mu \epsilon + \ldots \,.
\end{align}
The transformations in the metastable minimum become
\begin{align}
\delta_\epsilon \tilde \lambda^0 &= \epsilon^0  - \frac14 \alpha^2 \epsilon^0 + \ldots\,, \\
\delta_\epsilon \tilde \lambda^3 &= 0 - \alpha \epsilon^0  + \ldots \,, \\
\delta_\epsilon \tilde \psi &=(\bar{\tilde\lambda}^0\gamma^\mu \epsilon^0)\partial_\mu \tilde\psi -2\sqrt{2}e^\psi_{\underline\psi}( \alpha \bar{\tilde\lambda}^0\epsilon^0 + \frac14 \alpha^2 \bar{\tilde\lambda}^3 \epsilon^0) + \ldots\,,  \\
\delta_\epsilon \tilde A_\mu &= (\bar{\tilde\lambda}^0\gamma^\nu \epsilon^0) \tilde F_{\nu\mu} + 2\alpha \bar{\tilde\lambda}^0 \gamma_\mu \epsilon^0 + \frac12 \alpha^2 \bar{\tilde \lambda}^3 \gamma_\mu \epsilon^0 +  \ldots 
\end{align}
The first terms correspond to the standard non-linear transformations. Remember that the expansion of $\alpha$ around $\psi_{\rm min}$ is given by \eqref{eq:alphaexp}. We identify two types of corrections. First of all we observe that there are corrections that vanish in the probe limit $p/M \rightarrow 0$. These terms are just proportional to (the square of) $\psi_{\rm min} \sim p/M$ and reflect the shift towards the metastable minimum. In fact, if we could ignore the field $\delta\psi$ (as well as the spinor $\lambda^3$), the probe limit would consistently reproduce a subset of the non-linear supersymmetry transformations at the north pole. In other words, if the $\delta \psi$ and $\lambda^3$ fields were infinitely massive, the probe limit takes you to the north pole and a constrained superfield description of the goldstino and the gauge field would be adequate.

However, it can be seen from \eqref{eq:VDBIfluc} that the mass of the scalar $\psi$ is always of the same order of the potential energy scale in the metastable vacuum, so fluctuations in $\psi$ can never be decoupled. Interestingly the corrections that are proportional to $\delta \psi^2$ are all, except for the goldstino, proportional to $M/p$ suggesting that in the probe limit corrections become large and one should include (all) higher order terms. This is in line with the discussion of section \ref{sec:bosKPV}: at order $\delta \psi \sim p/M$ the quadratic approximation of the action breaks down. We are forced to conclude that a strict decoupling limit in which the effective description in terms of non-linearly realized supersymmetry becomes UV independent does not exist. As a consequence the validity of a  constrained superfield description is restricted. Just how restricted can be estimated by observing that the corrections become comparable to the shift term when the fluctuation $\delta \psi$ is of order $p/M$ or equivalently $\delta \psi \sim \psi_{\rm min}$. This should not come as a complete surprise, since this is where the expansion in $\delta \psi$ breaks down. We can translate this into a corresponding mass scale using the potential, giving a scale that is a factor of order $p/M$ smaller than the mass scale $m_\psi$, for any value of the string coupling. In other words, the description in terms of non-linearly realized supersymmetry seems to break down at scales far below the mass scale of relevance in the metastable vacuum.

Closing this section, we would like to make a final comment. It is important to realize that one should not perform an additional field redefinition at $\psi_{\rm min}$ that would remove the leading corrections. For instance, an additional field redefinition of $\lambda^3$ that removes the corrections at the same time modifies the form of the transformations at the north pole and also changes the fermionic mass matrix for $\lambda^0$. In this case, the redefined spinor cannot be identified with the massless goldstino.

\section{Comments and conclusions}\label{sec:conclusions}

Constrained superfields provide a powerful technique in the context of a universal (UV insensitive) low-energy description of spontaneously broken supersymmetry. A crucial requirement is a stable and large enough hierarchy between the scale of the fields that are projected out by the constraints and the relevant scale of the low energy effective theory. In some cases such a hierarchy might not be achievable, precluding the existence of a standard constrained superfield description. In general however the appropriate constrained superfield description is valid up to some energy scale that should be identified and compared to the supersymmetry breaking scale. In this work we studied the leading corrections to the nilpotent goldstino superfield description of anti-D3-branes in the GKP background from polarization effects. Our main observation is that the (non-linear) supersymmetry transformations in the metastable vacuum receive corrections that cannot be `decoupled' and actually become large in the probe limit $p \to 0$.

To arrive at that result we constructed, to leading order in the fields, the supersymmetric completion of the effective theory on an NS5-brane wrapped on an $S^2$ inside the transverse $S^3$ at the tip of the KS throat geometry of \cite{Kachru:2002gs}. We identified the massless goldstino of spontaneous supersymmetry breaking as well as the gauge field and transverse scalar $\psi$ that describes the position of the $S^2$ inside the $S^3$. In the absence of an orientifold plane that projects out the bosonic degrees of freedom, they should also transform non-linearly. In the metastable state we again identified $\lambda^0$, the singlet under the $SU(3)$ holonomy of the `internal' space, as the massless goldstino associated with the spontaneously broken supersymmetry. We argued this from the non-abelian point of view, which should hold provided we keep the radius of the $S^2$ small ($p/M \ll 1$). From the abelian perspective this should correspond to twisting the Dirac operator with a gauge field on the 2-sphere, as was done in \cite{Maldacena:2000yy}. A full treatment of the modified Dirac operator on the 4d reduced abelian NS5-brane should also reveal this zero mode at the position of the metastable minimum. We hope to come back to this question in future work.

We found that fluctuations of the $\psi$ scalar field around the metastable minimum cannot be decoupled. Moreover, corrections to the non-linear supersymmetry transformations become large at a scale far below the mass scale set by the scalar fluctuations in the metastable vacuum. The reduction factor is controlled by the ratio $p/M$ of the anti-brane number and the background flux. This limits a finite parameter window where an effective low-energy description of the metastable vacuum in terms of a constrained superfield is appropriate. 

This might not come as a total surprise. When the source of spontaneous supersymmetry breaking is intrinsically higher-dimensional, it might not admit any low-energy description in terms of (simple) constrained superfields. This is clearest for more energetic fluctuations around the metastable minimum, with $\delta \psi \sim p/M$. Those fluctuations are not localized around the metastable minimum, as they exceed the energy difference between the metastable state and the north pole (left maximum in figure \ref{fig:NS5pol}). However, they are still localized on the northern hemisphere of the $S^3$, as they have less energy than the absolute maximum of the potential. Those fluctuations describe full 6-dimensional fluctuations around the nilpotent superfield description of anti-D3 branes,  governing the non-linear transformations around the north pole $\psi = 0$. Increasing the scale of fluctuations even further will invalidate the non-linear description altogether, and will lead to a restoration of the linear transformations by higher-dimensional excitations.

The fluctuations we study in this paper are of a different nature. They capture excitations very close to the metastable minimum and obey $\delta \psi \ll p/M$. They can be captured in a four-dimensional language (albeit not with standard constrained superfields). Determining the relevant fluctuations in the KK reduction to four dimensions is subtle, since we discussed two different descriptions with opposite regimes of validity. The polarized NS5-brane point of view is only valid for a large $S^2$ and is hence intrinsically 6-dimensional. In section \ref{subsub:4dreduction}, we argued however from the dual non-abelian anti-D3 point of view that the set of lowest mass states of the KK spectrum in four-dimensions contains the massless goldstino. 

It is straightforward to check that the requirement $\delta \psi \ll p/M$ is a direct consequence of the relevance of higher order terms in the DBI action around the metastable vacuum. The expansion of the polarization potential around the metastable minimum \eqref{eq:VDBIfluc} shows that the higher order terms become important when $\delta\psi \sim \psi_{\rm min} \sim p/M$, as we also concluded from the supersymmetry transformations. From the low-energy effective field theory point of view the theory becomes strongly coupled as soon as $\delta\psi \sim \psi_{\rm min} \sim p/M$.

Our observations appear to be in line with the discussion of \cite{Dudas:2011kt}. The mass of the fluctuations $\delta \psi$ around the minimum of the potential is in fact of the same order as the supersymmetry breaking scale, as can easily be seen from \eqref{eq:VDBIfluc}
\be
 m_\psi^2 = \frac{4\sqrt{2}} {b_0^4}\sqrt{V_{\rm min}}\,.
\ee
As explained in \cite{Dudas:2011kt}, integrating out massive fields with masses of the order of the supersymmetry breaking scale does not lead to universal couplings of the goldstino and instead give rise to generalized holomorphic constraints on superfields. The UV dependence in our setup becomes apparent at scales $\delta\psi \sim \psi_{\rm min} \sim p/M$, where the cubic coupling starts to correct the supersymmetry transformations.  Whether and how this can be described in terms of generalized (higher order) constrained superfields, or in another approach such as the `goldstino brane' \cite{Bandos:2015xnf,Bandos:2016xyu}, is a question we hope to come back to in the future.

Let us finally briefly elaborate on what the general consequences of our findings might be in the context of string cosmology. Following the arguments of \cite{Dudas:2011kt}, to allow for a standard universal nilpotent superfield description one would require a stable hierarchy between the scale of supersymmetry breaking and the mass of the transverse scalar $\psi$. In the original KKLT scenario, the scale of supersymmetry breaking is set by the uplift energy of the metastable anti-D3 brane and hence seems to remain of the order of the mass of the $\psi$ fluctuations around the metastable vacuum. As a consequence the uplift with $p$ metastable polarized branes might lead to a similar breakdown of a  putative universal constrained superfield description at energies far below the supersymmetry breaking scale. An effective description of the metastable minimum by nilpotent superfields all the way up to the supersymmetry breaking scale with polarized anti-branes would require a version of the KKLT mechanism where the supersymmetry breaking scale and the uplift energy can be decoupled. Broad classes of such models are available: for instance in \cite{Kallosh:2004yh}, or anti-brane uplifts of an AdS minimum where supersymmetry is already broken, as in the Large Volume Scenario and related work \cite{Balasubramanian:2005zx,Conlon:2005ki,Aparicio:2015psl}. We hope to address some of these questions in future work.

\acknowledgments
We thank Daniel Baumann, Eric Bergshoeff, Nikolay Bobev, Ben Freivogel, Thomas Hertog, Dan Roberts, Gary Shiu, Hagen Triendl, Thomas Van Riet for discussions; Riccardo Argurio, Luca Martucci and Timm Wrase for feedback on a draft version of the paper; and Renata Kallosh and Timm Wrase for collaboration on related work.
BV thanks the Galileo Galilei Institute for Theoretical Physics for hospitality and the INFN for partial support during the completion of this work. BV was supported during the initial stages of work by: the European Commission through the Marie Curie Intra-European fellowship 328652--QM--sing and Starting Grant  of the European Research Council (ERC-2011-SrG 279617 TOI). Currently, BV is supported in part by the Interuniversity Attraction Poles Programme initiated by the Belgian Science Policy (P7/37), by the European Research Council grant no.\ ERC-2013-CoG 616732 HoloQosmos and the KU Leuven C1 grant Horizons in High-Energy Physics. This work is part of the Delta ITP consortium, a program of the Netherlands Organisation for Scientific Research (NWO) that is funded by the Dutch Ministry of Education, Culture and Science (OCW). The work of LA and JPvdS is also supported by the research program of the Foundation for Fundamental Research on Matter (FOM), which is part of the Netherlands Organization for Scientific Research (NWO).

\appendix{
\section{Details on fermions}\label{app:spinors}

In this appendix we review and apply the relevant details of the fermionic action of a D$p$-brane of \cite{Marolf:2003ye,Marolf:2003vf,Martucci:2005rb}, its supersymmetry transformations and gauge fixing.
We take the results for a D$5$ brane with worldvolume flux in the S-dual background to Klebanov-Strassler. We follow the conventions of \cite{Martucci:2005rb}. For easy comparison with the literature on gauge-fixed fermionic D-brane actions, we keep this appendix wholly in that `D5-frame' and we adapt notation slightly to match as much as possible the related work for D$p$-branes in flat space \cite{Bergshoeff:2013pia} used in the recent literature on
non-linear supersymmetries on anti-D3 branes  \cite{Kallosh:2014via,Bergshoeff:2015jxa,Vercnocke:2016fbt,Kallosh:2016aep}.

To transform the results of this appendix (`app') to the expressions used in the text, one has to apply the following S-duality rules to the NS5-frame:
\be
 H_3^{\rm app} = -F_3^{\rm text}\,,\qquad F_3^{\rm app}=  H_3^{\rm text}\,,\qquad e^{\Phi^{\rm app}} =  (g_s^{-1})^{\rm text}\,,\qquad {\cal F}^{\rm app} = 2 \pi g_s^{\rm text} {\cal F}^{\rm text}\,.
\ee

\subsection{Projection matrix}

We obtain the matrix $\Gamma_{\rm D5}$ from \cite{Martucci:2005rb}:
\be
\Gamma_{\rm D5} = -\begin{pmatrix} 0 & \betam\cr \betap&0 \end{pmatrix}\,,\label{eq:gammaD5}
\ee
with 
\be
\beta_\pm =\Gamma_{\rm D5}^{(0)} \frac{\sqrt{-\det g}}{\sqrt{-\det (g+{\cal F})}}\sum_k \frac {(\pm1)^k} {k! 2^k} \hat\Gamma^{\alpha_1 \ldots \alpha_{2k}} ({\cal F})_{\alpha_1 \alpha_2} \cdots({\cal F})_{\alpha_{2k-1} \alpha_{2k}}\,.
\ee
We have $\betap \betam = 1$ and the relation $\betam ({\cal F}) = \betap (- {\cal F})$. Note that hats on gamma matrices denote pull-backs on the worldvolume $\hat \Gamma_\alpha = \partial_\alpha X^M \Gamma_M$, and
\be
\Gamma^{(0)}_{\rm D5} = \frac {\varepsilon^{\alpha_1 \ldots \alpha_6} } {6!\sqrt{-\det g}} \hat \Gamma_{\alpha_1\ldots \alpha_6}\,.
\ee
We will split the field and the metric in a four-dimensional part (along the D3 worldvolume) and a transverse part along the two-sphere as:
\be
 {\cal F} = \ffour + \fsix\,,\qquad  ds^2 = ds^2_\parallel + ds^2_\perp\,.
\ee
It is not hard to see that the matrix in the projector splits as:
\be
\betap = \betap^{\perp} \betap^{\parallel}\,,\label{eq:gammaNS5check}
\ee
with
\begin{align}
\betap^{\perp} =&\frac {\varepsilon^{\alpha \beta} } {2!\sqrt{G_\perp}} \Gamma_{\alpha\beta} \frac{\sqrt{-\det G_\perp}}{\sqrt{-\det (G_\perp+\fsix)}}\left(1 + \frac 12 \fsix_{\alpha\beta} \Gamma^{\alpha\beta}\right)\,,\\
\betap^{\parallel} =& \frac { \varepsilon^{\mu_1 \ldots \mu_4}} {4!\sqrt{G_\parallel}} \hat \Gamma_{\mu_1\ldots \mu_4}  \frac{\sqrt{-\det G_\parallel}}{\sqrt{-\det (G_\parallel+\ffour)}}\left(1 + \frac 12 \ffour _{\mu_1\mu_2}\hat \Gamma^{\mu_1\mu_2}  + \frac 18 \ffour_{\mu_1\mu_2} \ffour_{\mu_3\mu_4} \hat \Gamma^{\mu_1 \mu_2\mu_3\mu_4} \right)
\end{align}
where Greek letters still refer to worldvolume indices, but we make a split: the middle of the alphabet to four dimensions ($\mu,\nu \ldots = 0,1,2,3)$ and the beginning to the two-sphere ($\alpha,\beta\ldots = 4,5$).

The calculation of the term $\betap^{\perp}$ follows straightforwardly from the discussion of section \ref{sec:bosKPV}, with
\be
\fsix = - Q(\psi) {\rm vol}_{S^2}\,.
\ee
The four-dimensional part of the projector parallels that of the projector dubbed $\beta$ in the appendix of \cite{Kallosh:2016aep}. Note that we only consider the bosonic terms, as fermionic terms in $\beta$ would take us beyond the quadratic fermionic order in the action. The result for $\betap^{\parallel}$ is
\begin{align}
 \betap^{\parallel} &= \Gamma_{\underline {0123}} \Big{(} 1  + \frac 12 \ffourtwo _{\mu\nu}\Gamma^{\mu\nu}+\partial^{\mu} X^I \Gamma_{I\mu}- \frac 12 g_{IJ}\partial_\mu X^I \partial^\mu X^J
 - \frac 14 \ffourtwo_{\mu\nu} \ffourtwo^{\mu\nu}+ \frac 14 \ffourtwo_{\mu\nu} (\star_4 \ffourtwo)^{\mu\nu} \Gamma_{\underline{0123}}\nonumber\\
 &\quad\,-\frac 12  \partial^{\mu} X^I \partial^{\nu} X^J\Gamma_{IJ\mu \nu} +\frac 12 \partial_{\rho} X^I  \ffourtwo _{\mu\nu}\Gamma^{I\rho\mu\nu} + \ldots\Big{)}\,.
\end{align}
The ellipses indicates terms higher order in fields and indices have been raised and lowered with the metric $G_\parallel$ and $X^I$ are the transverse coordinates. This is the straightforward covariantization of the kappa-symmetry matrix for a D3-brane.

Applied to one non-trivial transverse scalar $X^1 = \psi$, we have
 \begin{align}
 \betap^{\perp} &=\cos (\alpha) - \Gamma_{\underline{45}} \sin (\alpha)\\
 \betap^{\parallel} &= \Gamma_{\underline {0123}} \Big{(} 1  + \frac 12 \ffourtwo _{\mu\nu}\Gamma^{\mu\nu}+\partial^{\mu} \psi \Gamma_{\psi\mu}- \frac 12 g_{\psi\psi}\partial_\mu\psi\partial^\mu \psi
 - \frac 14 \ffourtwo_{\mu\nu} \ffourtwo^{\mu\nu}+ \frac 14 \ffourtwo_{\mu\nu} (\star_4 \ffourtwo)^{\mu\nu} \Gamma_{\underline{0123}}\nonumber\\
 &\quad\,+\frac 12 \partial_{\rho}\psi \ffourtwo _{\mu\nu}\Gamma^{\psi\rho\mu\nu} + \ldots\Big{)}\,.\label{eq:betas}
\end{align}

\subsection{Fermionic action}

We briefly describe how to get the mass terms of the action \ref{eq:action}. After gauge fixing $\theta_1=0$, and writing $\lambda = \theta_2$, he terms not involving derivatives define a mass matrix $\cal M$ as
\be
\bar \lambda {\cal M} \lambda \equiv \bar \lambda (1 - \Gamma_{\rm D5})[(\tilde M^{-1})^{\alpha\beta}\Gamma_\alpha W_\beta -\Delta ] \lambda.
\ee
We split the terms not involving a covariant derivative along the four-dimensions and the two-sphere as
\be
(\tilde M^{-1})^{\alpha\beta}\Gamma_\alpha W_\beta -\Delta = M_\parallel + M_\perp\,,
\ee
with
\begin{align}
M_\parallel &=[G_\parallel^{\mu\nu}\Gamma_\nu W_\mu -\Delta]\,,\\
M_\perp &= [(G_\perp + 2\pi g_s \sigma_3 {\cal F})^{-1}]^{\alpha\beta}\Gamma_\alpha W_\beta\,.
\end{align}
We find (using $\alpha,\beta$ for directions on the two sphere, and $\mu,\nu$ for four-dimensions)
\begin{align}
M_{\parallel} &= \frac 1 {8} \left(H_{\mu pq} \sigma_3 +  e^\Phi F_{\mu pq}\sigma_1\right)\Gamma^{\mu pq} - \frac 1 {24}\left( H_{mnp} \sigma_3 + e^\Phi F_{mnp} \sigma_1\right)\Gamma^{mnp}\\
 M_{\perp}&=  \sin^2(\alpha)\left( \frac 1 8 (\sigma_1F^{\alpha np}  + \sigma_3H^{\alpha np})\Gamma_{\alpha np}-\frac {1} {8} \sigma_1F^{mnp}\Gamma_{mnp}\right)\cr
&\quad\,+ \cos^2(\alpha)((\sigma_3 {\cal F})^{-1})^{\alpha\beta}\Bigg{(}\frac 18H_{\alpha pq}\sigma_3 \Gamma_{\beta}{}^{pq}
-\frac 1 {8\cdot 3!}e^{\Phi}F^{npq}\sigma_1 \Gamma_{\alpha\beta npq}\cr
&\quad\,+\frac 18 \left(2H_{\alpha \beta q} \sigma_3-e^{\Phi}F_{\alpha \beta q} \sigma_1\right)\Gamma^{q}
\Bigg{)}
\end{align}
with 
\be
\cos (\alpha) = -\frac{\sqrt{\det {\cal F}}}{\sqrt{-\det (G_\perp + {\cal F})}}\,,\qquad \sin (\alpha) = -\frac{\sqrt{\det G_\perp}}{\sqrt{-\det (G_\perp + {\cal F})}}\,.
\ee
The signs in these last two equations are chosen for later convenience.

Now we use that the flux $H_3$ is fully along $S^3$ and ${\cal F}$ is along $S^2$, while $F_3$ is orthogonal. So the non-zero terms in $M_{\parallel},M_{\perp}$ are
\begin{align}
M_{\parallel} &= - \frac 1 {24}\left( H_{mnp} \sigma_3 + e^\Phi F_{mnp} \sigma_1\right)\Gamma^{mnp}\\
 M_{\perp}
&= \sin^2(\alpha) \left( \frac 1 {12} (\sigma_3H^{m np})\Gamma_{m np}-\frac {1} {4\cdot 3!} \sigma_1e^\Phi F^{mnp}\Gamma_{mnp}\right)\cr
&\quad\,+ \cos^2 (\alpha)((\sigma_3 {\cal F})^{-1})^{\alpha\beta}\Bigg{(}-\frac 1 {8\cdot 3!}e^{\Phi}F^{npq}\sigma_1 \Gamma_{\alpha\beta npq}+\frac 14 H_{\alpha \beta q} \sigma_3\Gamma^{q}
\Bigg{)}\,,
\end{align}
which gives the result \eqref{eq:step1}.

From \eqref{eq:gammaD5} and \eqref{eq:betas} we find that for vanishing $F$ and neglecting the derivative terms on $\psi$ (as they are higher order in the action), we get
\be
\Gamma_{\rm D5} = -\begin{pmatrix} 0 & \betam\cr \betap&0 \end{pmatrix}\,, \qquad \beta_{\pm} = \Gamma_{\underline{0123}}(\pm \cos \alpha - \sin \alpha \Gamma_{\underline {45}})\,.\label{eq:gammaD5bis}
\ee
Now we use that for Majorana-Weyl bilinears only terms with three or seven gamma matrices are non-zero.
\be 
\bar{\lambda}\Gamma^{m_1\dots m_n} \lambda = 0 \qquad \text{for } n\notin \{3,7 \} 
\label{eq:MWcondition}
\ee
We now see that the last term in $M_\perp$ will not contribute at all and we find 
\begin{align}
\bar \lambda &(1-\Gamma_{\rm D5}){\cal M}  \lambda = \nonumber\\
& \frac 1 {24}\bar\lambda \left[\cos (2\alpha) H_{mnp} + \left(1 + \sin^2 (\alpha) - \frac 12 \cos^2(\alpha) ({\cal F}^{-1})^{\alpha\beta} \Gamma_{\alpha\beta} \right) e^\Phi \betap F_{mnp}\right] \Gamma^{mnp}\lambda\,.
\end{align}
With the identity $({\cal F}^{-1})^{\alpha\beta} \Gamma_{\alpha\beta} = 2 \tan(\alpha) \Gamma_{\underline{45}}$ and dropping again terms with the wrong number of $\Gamma$ matrices, we find 
\begin{align}
{\cal M} =  \frac 1 {24} \bar\lambda  \left(\cos (2\alpha) H_{mnp} + \cos( \alpha ) e^\Phi F_{mnp}\Gamma_{\underline{0123}}\right)\Gamma^{mnp}\lambda\,.
\end{align}
Finally we can use the Majorana-Weyl property $\Gamma_{(10)}\lambda =\lambda$, to write:
\be
F_{mnp}\Gamma_{\underline{0123}}\lambda = (\star_6 F)_{mnp}\lambda\,,
\ee
with $\star_6$ the Hodge star operator on the six-dimensional internal manifold. This yields the final result \eqref{eq:fermion_mass_matrix}:
\be
{\cal M} = \frac 1 {24}\left( \cos (2\alpha) H_{mnp} + e^\Phi \cos (\alpha)   F_{mnp}\Gamma_{\underline{0123}} \right)\Gamma^{mnp} \,.
\ee

\subsubsection{Fermionic action: orientifold compatible gauge choice}
For completeness, we show that taking the alternative gauge choice
\be
(1+\Gamma_{\rm D5})\theta=0 \quad \Leftrightarrow \quad \theta_1 = -\Gamma_{\underline{0123}}(\cos(\alpha) + \sin(\alpha) \Gamma_{\underline{45}}) \theta_2 \,,
\ee
to fix the kappa-symmetry we obtain the same mass matrix. This gauge choice is useful when one also wants to perform an orientifold projection, which has to be compatible with the gauge fixing condition. Using this condition, we can write the terms appearing in ${\cal M}_{\parallel}$ and ${\cal M}_{\perp}$ completely in terms of $\lambda\equiv \theta_2$.
\begin{align}
\bar{\theta}H_{mnp}\Gamma^{mnp}\theta &= 0 \nonumber \\
\bar{\theta}H_{mnp}\Gamma^{mnp}\sigma_3\theta &= -2\bar{\lambda}H_{mnp}\Gamma^{mnp}\lambda \nonumber\\
\bar{\theta}F_{mnp}\Gamma^{mnp}\sigma_1\theta &= -2\cos(\alpha)\bar{\lambda}F_{mnp}\Gamma^{mnp} \Gamma_{\underline{0123}}\lambda
\end{align}
We then find after some algebra that
\begin{align}
\bar{\lambda}{\cal M}_{\parallel}\lambda &= \frac{1}{12}\bar{\lambda}\left(H_{mnp} + e^\Phi \cos(\alpha) F_{mnp} \right)\Gamma^{mnp} \lambda \\
\bar{\lambda}{\cal M}_{\perp}\lambda &=-\frac{1}{6} \sin^2(\alpha)\, \bar{\lambda}  H_{mnp}\Gamma^{mnp} \lambda \, .
\end{align}
Where we again used \eqref{eq:MWcondition} to eliminate some terms. The total mass matrix is then given by
\be
\bar{\lambda}{\cal M} \lambda = \frac{1}{12} \bar{\lambda} \left( \cos(2\alpha) H_{mnp} + e^\Phi \cos(\alpha) F_{mnp} \Gamma_{\underline{0123}} \right) \Gamma^{mnp}\lambda \,
\ee
in agreement with the mass matrix in the gauge where $\theta_1=0$ up to a factor 2.

\subsection{Supersymmetry transformations}

The fields on the brane enjoy a combination of supersymmetry transformations, kappa-symmetry with spinorial parameter $\kappa$ and diffeomorphisms (we leave out the possibility of gauge transformations of the gauge field). To linear order in the fermions $\theta$, these are:
\begin{align}
\delta \theta &= \epsilon + [\unity +\Gamma_{\rm D5}] \kappa + \xi^\alpha \partial_\alpha \theta \,,\\
\delta X^m &=-\bar \theta \Gamma^m \epsilon+ \bar\theta \Gamma^m [\unity +\Gamma_{\rm D5}] \kappa+ \xi^\alpha \partial_\alpha X^m\,,\\
\delta A_\alpha &=-  \bar \theta \hat \Gamma_\alpha\sigma_3 \epsilon -  C_{\alpha m} \bar \theta \Gamma^m \epsilon + \bar  \theta  \hat \Gamma_\alpha \sigma_3[\unity +\Gamma_{\rm D5}]\kappa \nonumber\\
&\quad\, + C_{\alpha m} \bar \theta \Gamma^m [\unity +\Gamma_{\rm D5}] \theta+ \xi^\beta F_{\beta\alpha}\,.
\end{align}
As explained in \cite{Martucci:2005rb,Bergshoeff:2013pia}, we can fix the gauge redundancy in the following way. We fix kappa-symmetry by the spinor gauge choice $\theta_1 = 0$ or $(\unity + \sigma_3)\theta=0$, and requiring that this remains valid under the combined transformation 
\be
(\unity + \sigma_3)\delta \theta=0\,.
\ee
The diffeomorphism invariance can be fixed by requiring static gauge, such that $\delta X^\alpha = 0$. The background spinor obeys
\be
\epsilon_2 = \Gamma_{\underline{0123}} \epsilon_1\,.
\ee
This sets
\be
\epsilon^1 + \kappa^1 - \betam \kappa^2=0 \,,\qquad \text{and}\qquad \xi^\alpha = \bar\lambda \Gamma^\alpha [\unity  + \beta] \epsilon_2\,.
\ee
We will denote the transverse scalars by $X^I$ and with slight abuse of notation $\epsilon = -2\epsilon_2$. Then the SUSY transformations after fixing the kappa gauge that leave the quadratic action \eqref{eq:action_mass_matrix} invariant are (see also \cite{Martucci:2005rb})
\begin{align}
\delta_\epsilon \lambda &= -\frac 12 [\unity - \beta] \epsilon + {\cal O}(\lambda^2)\,,\\
\delta_\epsilon X^I &= \frac 12 \bar \lambda\Gamma^I [\unity  + \beta]\epsilon+ \xi^\alpha \partial_\alpha X^I + {\cal O}(\lambda^3)\,, \\
\delta_\epsilon A_\alpha &= -\frac 12 \bar \lambda (\hat \Gamma_\alpha + \Gamma_I \partial_\alpha X^I)[\unity  + \beta]  \epsilon + \frac 12 C_{\alpha m} \bar \lambda \Gamma^m [\unity  + \beta]  \epsilon+ \xi^\beta F_{\beta\alpha}+ {\cal O}(\lambda^3),
\end{align}
with
\begin{equation}
 \beta = \Gamma_{\underline{0123}} \betap\,, \qquad \xi^\alpha = -\frac12 \bar \lambda \Gamma^\alpha [\unity + \beta] \epsilon\,.
\end{equation}

}

\bibliographystyle{JHEP}
\bibliography{MetaStable-Constrained_JHEP}

\end{document}